\documentclass[11pt, a4paper]{article}
\pdfoutput=1

\usepackage[height=8.85in,width=6.75in]{geometry}

\usepackage{graphicx,rotating}     
\usepackage[bookmarksopen,colorlinks=true,linkcolor=light_blue,
citecolor=light_pink,urlcolor=dark_red,linktoc=all]{hyperref}

\usepackage{amsmath}
\usepackage{mathtools}
\usepackage{amsfonts}
\usepackage{amssymb}
\usepackage{slashed}
\usepackage{braket}
\usepackage{bm}
\usepackage{bbm}
\usepackage{cite}
\usepackage{comment}
\usepackage{mathrsfs}
\usepackage{xcolor}
\usepackage[utf8]{inputenc}
\usepackage[footnotesize]{caption}
\usepackage{bbold}
\usepackage{xfrac}
\usepackage{physics}

\usepackage[lf]{Baskervaldx} 
\usepackage[bigdelims,vvarbb]{newtxmath} 
\usepackage[cal=boondoxo]{mathalfa} 

\newcommand{\Mpl}{M_{\text{Pl}}}
\newcommand{\diff}{D_\text{s}}

\makeatletter
\@addtoreset{equation}{section}

\makeatother

\usepackage{multicol}
\definecolor{dark_red}{rgb}{0.7, 0., 0.}
\definecolor{light_pink}{rgb}{1,0.4,0.4}
\definecolor{light_blue}{rgb}{0.284602,0.317763,0.963947}
\definecolor{cred}{RGB}{180,50,40}
\definecolor{darkgreen}{RGB}{0, 100, 0}
\definecolor{desy_blue}{HTML}{009EE2}
\definecolor{desy_orange}{HTML}{FD8800}
\definecolor{forestgreen}{HTML}{228B22}
\definecolor{ochre}{HTML}{CCAA2B}

\newcommand{\GeV}{\,\mathrm{GeV}}

\begin{document}

\hypersetup{pageanchor=false}
\begin{titlepage}

\begin{center}

\hfill KEK-TH-2602\\
\hfill TU-1224\\

\vskip 0.5in

{\Huge \bfseries
Perturbative reheating and thermalization\\of pure Yang-Mills plasma\\
}
\vskip .8in

{\Large Kyohei Mukaida$^{\circ,\triangle}$, Masaki Yamada$^{\bullet,\blacktriangle}$}

\vskip .3in
\begin{tabular}{ll}
$^\circ$& \!\!\!\!\!\emph{Theory Center, IPNS, KEK, 1-1 Oho, Tsukuba, Ibaraki 305-0801, Japan}\\
$^\triangle$& \!\!\!\!\!\emph{Graduate University for Advanced Studies (Sokendai), }\\[-.3em]
& \!\!\!\!\!\emph{1-1 Oho, Tsukuba, Ibaraki 305-0801, Japan}\\
$^\bullet$& \!\!\!\!\!\emph{FRIS, Tohoku University, Sendai, Miyagi 980-8578, Japan}\\
$^\blacktriangle$& \!\!\!\!\!\emph{Department of Physics, Tohoku University, Sendai, Miyagi 980-8578, Japan}\\
\end{tabular}

\end{center}
\vskip .6in

\begin{abstract}
\noindent
We investigate the thermalization of high-energy particles injected from the perturbative decay of inflaton during the pre-thermal phase of reheating in detail. In general, thermalization takes a relatively long time in a low-temperature plasma; therefore, the instantaneous thermalization approximation is not justified, even for the reheating of the Standard Model (SM) sector. We consider a pure Yang--Mills (YM) theory as an approximation of the SM sector or a possible dark sector, considering the Landau--Pomeranchuk--Migdal effect, a quantum interference effect in a finite temperature plasma. We perform the first numerical calculation to solve the time evolution of the system, including the redshift due to the expansion of the Universe, and show the details of the temperature evolution near the maximum and the behavior of the quasi-attractors at later times. The maximal temperature $T_\text{max}$ and time scale $t_\text{max}$ are determined quantitatively, such as $T_\text{max} \simeq 0.05 \times (\Gamma_I \Mpl^2/m_I^3)^{2/5} m_I$ and $t_\text{max} \simeq 2 \times 10^3 \times (\Gamma_I \Mpl^2/m_I^3)^{-3/5} m_I^{-1}$ in the SM-like system, where $m_I$ and $\Gamma_I$ are the mass and decay rate of inflaton. We also provide a similar formula for pure $\operatorname*{SU}(N)$ and $\operatorname*{SO}(N)$ YM theories for general values of $N$ and coupling constant $\alpha$, including $T_\text{max} \propto \alpha^{4/5}$ and $t_\text{max} \propto N^{-2} \alpha^{-16/5}$ behaviors and their numerical coefficients. The thermalization occurs in a finite time scale, resulting in a lower maximal temperature of the Universe after inflation than that under the instantaneous thermalization approximation.
\end{abstract}

\end{titlepage}

\tableofcontents
\thispagestyle{empty}
\renewcommand{\thepage}{\arabic{page}}
\renewcommand{\thefootnote}{$\natural$\arabic{footnote}}
\setcounter{footnote}{0}
\newpage
\hypersetup{pageanchor=true}

\section{Introduction}
\label{sec:intro}

The Big Bang cosmology has been confirmed by the success of the Big Bang nucleosynthesis (BBN) theory~\cite{Schramm:1997vs,Steigman:2007xt} and the observation of cosmic microwave background (CMB)~\cite{Penzias:1965wn,Mather:1993ij}.
After the Big Bang, the temperature of the background plasma decreases via the redshift due to the expansion of the Universe.
The Universe might experience several phase transitions during cosmological thermal history, including the electroweak and QCD phase transitions.
Several models of physics beyond the Standard Model (SM) also predict other cosmological phase transitions.
In general, phase transitions lead to rich phenomenology in cosmology, such as the emission of gravitational waves (GWs)~\cite{Kosowsky:1992rz,Kosowsky:1991ua,Kosowsky:1992vn,Kamionkowski:1993fg} and formation of topological defects~\cite{Vilenkin:1984ib}.
The dynamics of topological defects also lead to the emission of GWs~\cite{Vilenkin:1981bx,Vilenkin:1984ib,Accetta:1988bg,Caldwell:1991jj,Gouttenoire:2019kij,Vilenkin:1981zs,Preskill:1991kd,Saikawa:2017hiv,Vilenkin:1982hm,Leblond:2009fq,Buchmuller:2019gfy,Chang:1998tb,Dunsky:2021tih}.
These GW signals are outstanding signatures of physics beyond the SM that is not accessible by collider experiments, and therefore, understanding the detailed thermal history is important to reveal the physics beyond the SM.
For this purpose, the maximal temperature of the Universe is an important quantity.
If the maximal temperature is higher than the critical temperature of a phase transition, one can expect that the Universe experiences the phase transition as it cools down due to the redshift.

The several problems of Big Bang cosmology, related to its initial condition, can be addressed by inflation~\cite{Guth:1980zm,Starobinsky:1980te,Sato:1980yn}.
After inflation, the Universe is reheated by the decay of inflaton into radiation. The temperature of radiation reaches its maximum at a specific time and then cools down due to the redshift.
One may assume \textit{instantaneous reheating}, in which the radiation-dominated era is followed by inflation without an inflaton-dominated era.
However, this scenario is considerably simplified in several cases.
If the inflaton couples to radiation without any suppression, its decay is considerably fast and might cause a non-perturbative process called preheating~\cite{Kofman:1994rk,Kofman:1997yn}.
On the contrary, if the inflaton couples to radiation feebly, then the inflaton tends to dominate the Universe for a while and then slowly decays into radiation perturbatively.
In this paper, we focus on the latter scenario.

The slow decay of the inflaton into radiation does not imply that the radiation is thermalized instantaneously after production.
Namely, \textit{instantaneous thermalization} is not generally justified and the thermal history during the \textit{pre-thermal phase} is not trivial.
Let us specifically consider a suppressed decay of inflation into gluons, \textit{e.g.,} via a (Planck-suppressed) higher-dimensional operator.
For a typical inflation model, the inflaton mass is very large.\footnote{
  Although the decay rate is sufficiently small to justify the perturbative analysis, the temperature of plasma can be greater than the inflaton mass. In this case, the decay rate gets modified by the thermal effects~\cite{Mukaida:2012bz,Mukaida:2012qn,Drewes:2013iaa}.
}
The decay of such a heavy particle produces very high energy particles, such as gluons, in background plasma.
The thermalization of high--energy gluons is quite non-trivial
because they should lose significant amount of energy to be thermalized and absorbed into thermal plasma~\cite{Baier:2000sb,Kurkela:2011ti,Kurkela:2014tea,Fu:2021jhl}.
In fact, our previous studies have shown that the thermalization-time scale is generally significantly longer than the Hubble-time scale in the early stage of reheating, and the maximal temperature of the Universe may be significantly suppressed~\cite{Harigaya:2013vwa,Mukaida:2015ria,Passaglia:2021upk,Drees:2021lbm,Drees:2022vvn,Mukaida:2022bbo}.
Detailed thermalization also provides a novel dark matter (DM) production mechanism in the pre-thermal phase~\cite{Harigaya:2014waa,Harigaya:2019tzu,Garcia:2018wtq,Mukaida:2022bbo}.
Several previous studies have focused on the DM production, in which case the particle distribution is reduced to a stationary solution, and the analysis can be significantly simplified.

The bottleneck process for thermalization is the splitting of a high-energy gluon into low-energy daughter particles.
The splitting process experiences an interference between wavepackets, known as the Landau--Pomeranchuk--Migdal (LPM) effect~\cite{Landau:1953um,Migdal:1956tc,Gyulassy:1993hr,Arnold:2001ba,Arnold:2001ms,Arnold:2002ja}, and therefore has a strongly suppressed rate of $\Gamma_{\rm split} = \alpha^2 \sqrt{T^3/p}$, where $p$ is the energy of the parent particle, and $T$ is the (effective) temperature of the background plasma.\footnote{
This splitting rate is for the cases with non-Abelian interactions, which we focus on in this paper. The splitting rate for Abelian interactions is different; however, a qualitatively similar cosmological thermal history is obtained~\cite{Harigaya:2014waa,Mukaida:2022bbo}.
}
Comparing this with the Hubble expansion rate,
the maximal temperature of the Universe can be estimated as $T_\text{max} \sim \alpha^{4/5} (\Gamma_I \Mpl^2/m_I^3)^{2/5} m_I$, where $\Mpl$ is the reduced Planck mass, $\Gamma_I$ is the inflaton decay rate, and $m_I$ is the inflaton mass~\cite{Harigaya:2013vwa,Mukaida:2015ria}.
This can be several orders of magnitude smaller than the naive result under the {\it instantaneous thermalization} approximation.
This qualitative discussion demonstrates the importance of a detailed investigation of thermalization for inflaton-decay products to pin down the maximal temperature of the Universe.

In this paper, we numerically solve the detailed Boltzmann equation to calculate the time dependence of temperature during the pre-thermal phase, considering the LPM effect.
To minimize the numerical cost, a pure Yang--Mills (YM) theory is considered. This is a good approximation for the SM sector and also motivated by a thermalization of the dark sector that may explain DM.
We provide a quantitative formula for the maximal temperature and thermalization time scale.
This is the first work on the quantitative time evolution of thermal plasma during the pre-thermal phase.

The organization of this paper is as follows.
In Sec.~\ref{sec:pre}, we briefly review the thermalization process of a high-energy particle under the cosmological expansion.
Moreover, a qualitative estimation of thermal history is explained.
In Sec.~\ref{sec:numerical}, we explain our numerical method and show our results. The results are consistent with the qualitative discussion, but provide more information, including numerical prefactors.
We consider a model that mimics the SM sector and a model of the pure YM dark sector.
In the latter case, the gauge group $G$ is assumed to be $\operatorname*{SU}(N)$ and $\operatorname*{SO}(N)$.
A formula for maximal temperature is also shown in a large $N$ and small gauge coupling limit.
Section~\ref{sec:conclusions} is devoted to discussion and conclusions.

\section{Kinetic equations}
\label{sec:pre}

\subsection{Warmup}
\label{sec:setup}

Let us start our discussion by neglecting the thermalization of the produced particles as a warmup.
In this case, the relevant Boltzmann equation is given as follows:
\begin{equation}
  \qty( \frac{\partial }{\partial t} - H p \frac{\partial }{\partial p} ) f(p,t)= \mathcal{S},
  \label{eq:Boltzmann_warmup}
\end{equation}
where $p$ is the physical momentum, and
$H$ is the Hubble parameter.
The distribution function of gluon per one degree of freedom is denoted by $f$, that is, the total number density of gluon is obtained by multiplying the degrees of freedom.
The source term is (approximately) given by a delta function at $p = p_0$.
Throughout this paper, we consider the case where the primary particles originate from
the two-body decay of a heavy particle, \textit{e.g.,} inflaton, with number density $n_I (t)$, mass $m_I$, and decay rate $\Gamma_I$.
The source term is expressed as follows:
\begin{align}
&\mathcal{S} =
\frac{1}{\nu_g}
\frac{\dd \Gamma_I}{\dd p} \frac{2\pi^2}{p^2} n_I(t),
\\
&\frac{\dd \Gamma_I}{\dd p}= 2 \Gamma_I \delta \qty( p - p_0 ), \qquad
p_0 = m_I / 2,
  \label{eq:source}
\end{align}
where $\nu_g$ is the degrees of freedom of the gluon.
If the heavy particle behaves as pressureless matter, we have
\begin{equation}
  n_I(t) = n_I (t_0) \qty[ \frac{a(t_0)}{a(t)} ]^3 e^{-\Gamma_I t} ,
  \label{eq:nI}
\end{equation}
where $a(t)$ is the scale factor, and $t_0$ is the reference time.

Eq.~\eqref{eq:Boltzmann_warmup} can be solved, such as
\begin{align}
  &f(p,t) = f_\text{h}(p,t) \equiv
  f_\text{h}(p_0)
  \qty[ \frac{a_0 p_0}{a(t) p} ]^{3-1/n}
  e^{- \Gamma_I t_0 \qty[\frac{a(t) p}{a_0 p_0} ]^{1/n}}
  \theta \qty(p_0 - p )
  \theta \qty(p  - \frac{a(t_\text{inf})}{a(t)} p_0),
  \label{eq:initialf}
\\
  &f_\text{h} (p_0) \equiv 6 \pi^2 \nu_g^{-1} \qty[ \frac{2 p_0 n_I (t_0) }{3 H_0^2 \Mpl^2} ]
  \qty( \frac{H_0 \Gamma_I \Mpl^2}{p_0^4} ),
\end{align}
where
$H = n/t$ and $a(t)/a_0 = (t/t_0)^{n}$.
In this paper, we focus on the thermalization in the matter-dominated epoch, in which $n = 2/3$.
Hereafter, we neglect the exponential factor in \eqref{eq:initialf} because
we are primarily interested in the thermalization that occurs in the regime of $\Gamma_I < H(t)$.
The momentum $a(t_{\rm inf})p_0/a(t) $ in the second Heaviside theta function in \eqref{eq:initialf} represents
the redshifted momentum of gluons
generated at the end of inflation $t_\text{inf}$.
We take $a(t_\text{inf}) \to 0$ for simplicity throughout this paper because its precise value does not affect our result qualitatively.

\subsection{Qualitative discussion}
\label{sec:qual}

We move on to the discussion on the thermalization of pure YM plasma produced by the decay of a heavy particle, \textit{e.g.,} inflaton.
Before presenting the concrete kinetic equations, we briefly summarize the basic assumptions and qualitative discussion of reheating and thermalization. See also Ref.~\cite{Mukaida:2015ria}.
As emphasized in the introduction, we focus on the case in which the decay rate of inflaton, $\Gamma_I$, is sufficiently small such that the reheating can be well approximated by the perturbative inflaton decay.
It is convenient to use the following combination for the decay rate $\Gamma_I$:
\begin{equation}
  \frac{\Gamma_I}{m_I^3/\Mpl^2}
  \left( = \frac{\nu_g p_0 t_0 f_\text{h}(p_0)}{32 \pi^2} \right),
  \label{eq:gamma-to-fh}
\end{equation}
which is assumed to be smaller than unity.
This is typically expected at least when the inflaton decay rate is smaller than that induced by the dimension-five Planck-suppressed operators, such as $\kappa_{5} I G^a_{\mu\nu} G^{a \mu\nu} / \Mpl$ and $\kappa_{5} I G^a_{\mu\nu} \tilde G^{a \mu\nu} / \Mpl$ with $\kappa_5 \ll 1$, where $I$ represents the inflaton.
In this case, the reheating temperature is significantly smaller than the inflaton mass.
Hence, we generically expect that the temperature of the ``dilute'' plasma, which already exists before the completion of reheating, is smaller than the typical energy of gluons just after their production, at least in the later stage of pre-thermal phase.

The bottleneck process of thermalization in such cases is given by the splittings of high-energy gluons into the background plasma.
In general, a splitter of momentum $p$ can emit a splittee at a scale of $k < p$ only after $t \geqslant k / k_\perp^2$ with $k_\perp$ being a momentum of $k$ transverse to the direction of $p$ because otherwise, quantum mechanical interference between the splitter and splittee prevents splittee formation.
Suppose there exist soft-thermalized populations of gluons with a temperature $T_\text{s}$, whose condition is specified later.
The interactions with the soft thermal plasma induce random diffusions of the transverse momentum, which is estimated as $k_\perp^2 \sim \hat q t \sim D_s \alpha^2 T_\text{s}^3 t$.\footnote{
  Here, we neglect the running of $\alpha$ for simplicity. See the later discussion on how the running modifies the estimation.
}
Here, we have included the factor $\diff$ that depends on the degrees of freedom responsible for the transverse diffusion.
One may estimate its dependence as
\begin{equation}
  \diff \sim C_\text{A} \sum_{i} \frac{\nu_i}{d_i} t_i,
\end{equation}
where the summation is taken over species $i$ contributing to the transverse diffusion, the degrees of freedom for $i$ are denoted by $\nu_i$, the dimension and normalization of representation for $i$ are $d_i$ and $t_i$, respectively, and the quadratic Casimir for the adjoint representation is $C_\text{A}$.
For pure YM plasma, $\diff \sim C_\text{A}^2$.
Combining these two estimations, we find the formation momentum below which the splittees can be emitted for a given $t$
\begin{equation}
  k < k_\text{form} (t) \equiv \diff \alpha^2 T_\text{s}^3 t^2 \,.
\end{equation}
Equivalently, the formation time before which the splittees of momentum $k$ cannot be emitted is given as follows:
\begin{equation}
  t > t_\text{form}(k) \equiv \sqrt{\frac{k}{\hat q}} \sim \qty(\alpha T_\text{s})^{-1} \sqrt{\frac{k}{\diff T_\text{s}}}.
  \label{eq:tform}
\end{equation}
Once this condition is met, the splittee of momentum $k$ can be formed with a probability of $\alpha$, which leads to the LPM-suppressed splitting rate, $\Gamma_\text{LPM} (k) \sim C_\text{A} \alpha / t_\text{form} (k)$.
On the other hand, for $k > k_\text{form} (t)$, the interactions with the medium are not sufficient to build up the transverse momentum, whereas a large-angle emission of $\theta^2_\text{vac} \geqslant 1/(k t)$ is always allowed quantum-mechanically with a probability of $\alpha$ (up to a logarithmic factor), that is, Dokshitzer--Gribov--Lipatov--Altarelli--Parisi (DGLAP) vacuum shower.
Hence, the splitting rate, where a splitter of momentum $p$ splits into daughters of $k$ and $(p - k)$, can be obtained as (see also Ref.~\cite{Kurkela:2014tla})
\begin{equation}
  \Gamma_\text{split} (k) \sim
  \begin{cases}
    \Gamma_\text{LPM} (k) \sim C_\text{A} \alpha^2 T_\text{s} \sqrt{\frac{\diff T_\text{s}}{k}} &\text{for} \quad T_\text{s} < k < k_\text{form} (t),\\
    \Gamma_\text{DGLAP} (k) \sim C_\text{A} \frac{\alpha}{t} &\text{for} \quad k_\text{form} (t) < k < \frac{m_I}{2},
  \end{cases}
  \label{eq:splitting_qual}
\end{equation}
where logarithmic factors are dropped for simplicity.

The emitted splittees of momentum $k$ immediately cascades and participate in the soft thermal plasma within the Hubble time if $\Gamma_\text{LPM} (k) > H \sim 1/t$, which reads
\begin{equation}
  k < k_\text{split} (t) \equiv C_\text{A}^2 \alpha^2 k_\text{form} (t) = C_\text{A}^2 \diff \alpha^4 T_\text{s}^3 t^2.
  \label{eq:ksplit}
\end{equation}
The energy conservation implies $\rho_\text{s} (t) \sim k_\text{split} (t)\, \Gamma_I t\, n_I (t)$ with $\rho_\text{s}$ being the energy density of the soft sector.
Assuming the thermalization of the soft sector, \textit{i.e.}, $\rho_\text{s} \sim g_\star T_\text{s}^4$ with $g_\star$ being the effective relativistic degrees of freedom in the soft thermal plasma, \textit{e.g.,} $g_\star = \nu_g = 2 d_\text{A}$ for pure YM plasma, we obtain
\begin{equation}
  \frac{T_\text{s} (t)}{m_I} \sim g_\star^{-1} C_\text{A}^2 \diff \alpha^4 \qty( \frac{\Gamma_I}{m_I^3 / \Mpl^2} )\, m_I t.
  \label{eq:Tsoft}
\end{equation}
Now, the condition for the soft-sector thermalization can be derived.
To maintain the thermal distribution, the large-angle scatterings among the soft populations should be much faster than the cosmic expansion $\diff \alpha^2 T_\text{s} (t) > H$,%
\footnote{
Here we write the dependence of degrees of freedom for the large-angle scattering rate as $D_s$ for notational simplicity. Although this is correct at least for pure YM plasma, it is different from $D_s$ for a general case.
}
which leads to
\begin{equation}
  t > t_\text{soft} \equiv g_\star^{1/2} C_\text{A}^{-1} \diff^{-1} \alpha^{-3} \qty( \frac{\Gamma_I}{m_I^3 / \Mpl^2} )^{-1/2} m_I^{-1}.
  \label{eq:tsoft}
\end{equation}
Throughout this paper, we restrict ourselves to $t \gg t_\text{soft}$.

With time, $k_\text{split}$ grows continuously, and eventually $k_\text{split} (t) >  m_I$, or equivalently $\Gamma_\text{split} (m_I) > H$, which occurs at
\begin{equation}
  t > t_\text{max} \equiv
  g_\star^{3/5} C_\text{A}^{-8/5} \diff^{-4/5} \alpha^{-16 / 5} \qty(\frac{\Gamma_I}{m_I^3 / \Mpl^2})^{-3/5} m_I^{-1}.
  \label{eq:tthermal}
\end{equation}
Subsequently, the gluons generated by the decay of inflaton immediately break up into the soft thermal plasma.
The temperature of the soft sector is then given by $\rho_\text{s}\sim m_I \Gamma_I t n_I$, which yields
\begin{equation}
  \frac{T_\text{s} (t)}{m_I} \sim g_\star^{-1/4} \qty(\frac{\Gamma_I}{m_I^3 / \Mpl^2})^{1/4} \qty(m_I t)^{-1/4},
\end{equation}
for $t_\text{max} < t < \Gamma_I^{-1}$.
The reheating is completed at $t \sim \Gamma_I^{-1}$, and the temperature at that time is often referred to as the \textit{reheating temperature}, $T_\text{R} \sim g_\star^{-1/4} \sqrt{\Mpl \Gamma_I}$.

To sum up, in addition to the hard distribution given in Eq.~\eqref{eq:initialf}, the splittings of high-energy gluons yield the cascades towards the soft thermal plasma.
Its distribution can be estimated as
\begin{alignat}{2}
  \nu_g f_\text{s} (k) &\sim
  \begin{cases}
    \frac{\nu_g T_\text{s} (t)}{k} &\cdots \quad k < T_\text{s} (t) \\
    g_\star
        \frac{\diff^{1/2}}{C_\text{A}}
    \Big(\frac{t_\text{soft}}{t} \Big)^2 \Big(\frac{T_\text{s} (t)}{k} \Big)^{7/2} &\cdots \quad T_\text{s} (t) < k < k_\text{form} (t) \\
    C_\text{A} \alpha \Big(\frac{1}{m_I t} \Big) \Big(\frac{\Gamma_I}{m_I^3 / \Mpl^2} \Big) \Big( \frac{m_I}{k} \Big)^3 &\cdots \quad k_\text{form} (t) < k < m_I/2
  \end{cases}
  && \qquad \text{for} \quad t_\text{soft} < t < t_\text{max},
  \label{eq:distribution1}
  \\
  \nu_g f_\text{s} (k) &\sim
  \begin{cases}
    \frac{\nu_g T_\text{s} (t)}{k} &\cdots \quad k < T_\text{s} (t) \\
    \frac{\Gamma_I}{\Gamma_\text{LPM}(m_I)} \frac{n_I (t)}{m_I^3}\Big(\frac{m_I}{k}\Big)^{7/2}  &\cdots \quad T_\text{s} (t) < k < m_I/2
  \end{cases}
  &&\qquad \text{for} \quad  t_\text{max} < t < \Gamma_I^{-1}.
  \label{eq:distribution2}
\end{alignat}
Note that the hard population given in Eq.~\eqref{eq:initialf} vanishes for $t > t_\text{max}$.
The temperature of the soft sector is given by
\begin{equation}
  \frac{T_\text{s} (t)}{m_I} \sim \begin{cases}
    g_\star^{-1} C_\text{A}^2 \diff \alpha^4 \Big(\frac{\Gamma_I}{m_I^3 / \Mpl^2} \Big)\, m_I t &\text{for} \quad t_\text{soft} < t < t_\text{max}, \\[.5em]
    g_\star^{-1/4} \Big(\frac{\Gamma_I}{m_I^3 / \Mpl^2} \Big)^{1/4} \qty( m_I t )^{-1/4} &\text{for} \quad t_\text{max} < t < \Gamma_I^{-1},
  \end{cases}
  \label{eq:Ts}
\end{equation}
which is maximized at $t \sim t_\text{max}$ as
\begin{equation}
  T_\text{max} \sim g_\star^{-2/5} C_\text{A}^{2/5} \diff^{1/5} \alpha^{4/5} \qty(\frac{\Gamma_I}{m_I^3 / \Mpl^2})^{2/5} m_I.
  \label{eq:Tmax}
\end{equation}
The rest of this paper is devoted to refining this qualitative understanding through numerical simulations.

\subsection{In-medium splitting function}
\label{sec:kineq}

The kinetic equation that describes the reheating and thermalization via the production of high-energy gluons after inflation is summarized as follows~\cite{Arnold:2002zm,Mukaida:2022bbo}:
\begin{equation}
  \qty( \frac{\partial }{\partial t} - H p \frac{\partial }{\partial p} ) f(p,t)= \mathcal{S}+ \mathcal{C}_{1 \leftrightarrow 2} [f] + \mathcal{C}_{2 \leftrightarrow 2} [f],
  \label{eq:Boltzmann0}
\end{equation}
where the splittings of high-energy gluons are described by $\mathcal{C}_{1 \leftrightarrow 2}$, and
the elastic scatterings are represented by $\mathcal{C}_{2 \leftrightarrow 2}$, which lead to the thermalization of the soft sector.
Throughout this paper, we focus on the time scale much longer than the elastic scatterings, \textit{i.e.,} $t \gg t_\text{soft}$.
Hence, we simply assume that the splittees of momentum $k$ immediately get thermalized as long as $k_\text{IR} > k$ and impose the thermalization by hand without explicitly including $\mathcal{C}_{2 \leftrightarrow 2}$ in the numerical simulations.
The results of our numerical simulations should be insensitive to the choice of $k_\text{IR}$ as long as $T_\text{s} (t) \lesssim k_\text{IR} \ll k_\text{split} (t)$.

Throughout this paper, we consider the case where the cascades of high-energy gluons are dominated by the splittings into low-energy gluons.
This restriction is trivially fulfilled for weakly coupled pure YM theories and holds for the high-energy gluons in the SM.
The splitting of gluons is encoded in $\mathcal{C}_{1 \leftrightarrow 2}$ as
\begin{equation}
  \mathcal{C}_{1 \leftrightarrow 2} [f]
  =
  \frac{(2\pi)^3}{p^2 \nu_g}  \qty[ -
  \int_0^p  \dd k \,
    \gamma_{g \leftrightarrow g g} \bigl(p; k, p-k \bigr)
    f (p)
    +
  \int_0^\infty \dd k \,
    2 \gamma_{g \leftrightarrow gg} \bigl(p+k; p, k \bigr) f (p+k)
  ].
  \label{eq:C12}
\end{equation}
The splitting function of gluons in the presence of thermal plasma is
\begin{equation}
  \gamma_{g\leftrightarrow g g}(P; xP, (1-x)P)
  = \frac{1}{2} \frac{d_{\text{A}} C_{\text{A}} \alpha}{(2\pi)^4 \sqrt2}
  \,
  \frac{P^\text{(vac)}_{g \leftrightarrow gg} (x)}{x(1-x)}
  \,
  \mu_{\perp}^2(P; 1,x,1-x) , \qquad
  P^\text{(vac)}_{g \leftrightarrow gg} (x) \equiv \frac{1^4+x^4+(1-x)^4}{x (1-x)},
\label{eq:gamma_ggg}
\end{equation}
where the fine structure constant $\alpha$ is evaluated at a scale $P$, the well-known DGLAP splitting function for gluons is denoted as $P^\text{(vac)}_{g \leftrightarrow gg} (x)$, the degree of freedom is $\nu_g = 2 d_\text{A}$, the dimension of the adjoint representation is $d_{\text{A}}$, and its quadratic Casimir is $C_{\text{A}}$;
for $\operatorname{SU}(N)$ and $\operatorname{SO}(N)$ gauge theories, $(d_\text{A}, C_\text{A}) = (N^2-1,N)$ and $(N(N-1)/2, 2N-4)$, respectively.
The transverse momentum squared at the formation time is denoted by $\mu_\perp^2$, which is developed by interactions with the soft thermal plasma of gluons.
It can be computed by solving a self-consistent equation, as done in Refs.~\cite{Arnold:2008zu}, whose result at the leading log is given by
\begin{align}
  \mu^4_\perp \qty(P; x_1, x_2, x_3) =
      \frac{2}{\pi} \,
      x_1 x_2 x_3 \,
      P\,
    \frac{ \alpha \qty(m_\text{D}) - \alpha \qty(Q_{\perp}) }{ - b_\alpha / \qty( 64 \pi^3 )}\, \mathcal{N}
    \frac{C_A}{2} \qty( x_1^2 + x_2^2 + x_3^2 ),
    \label{eq:mu_perp}
\end{align}
with $b_\alpha$ being the one-loop $\beta$-function coefficient, \textit{e.g.,} $b_\alpha = - 11 C_\text{A} / 3$ in the pure YM theory.
The factor $\mathcal{N}$ is proportional to the number density responsible for the transverse momentum diffusions and is given as follows:
\begin{align}
  \mathcal{N}
  &\equiv
  \sum_i
  \frac{\nu_i}{d_{i}} t_{i}
  \int \frac{\dd^3\ell}{(2\pi)^3} \,  f_\text{s}(\ell)
  \\
  &= 2 C_\text{A}  \frac{\zeta (3)}{\pi^2} T^3_\text{s} \qquad \text{for} \quad \text{pure YM},
  \label{eq:N_a}
\end{align}
where the summation is taken over species $i$ contributing to the transverse diffusion, the distribution function of the soft population is denoted as $f_\text{s}$, and the Riemann zeta function at $3$ is $\zeta (3) \simeq 1.20206$.
Note that the transverse momentum diffusion is dominated by the soft thermal plasma for $t \gg t_\text{soft}$.
In the second line, we used the fact that the normalization of the adjoint representation fulfills $t_\text{A} = C_\text{A}$.
We also define the following quantities:
\begin{align}
  \qty( \frac{Q_{\perp}}{m_{D}} )^2
  &\sim \left(\frac{P}{T_\text{s}}\right)^{1/2} \ln^{1/2}\left(\frac{P}{T_\text{s}}\right) ,
  \\
  m_\text{D}^2 &=
    8\pi \alpha
    \sum_i
    \frac{\nu_i}{d_i} t_i \int\frac{\dd^3 \ell}{(2\pi)^3} \frac{f_\text{s}(\ell)}{\ell}
    \\
    &= 4 \pi \alpha (T_\text{s}) \frac{C_\text{A}}{3} T_\text{s}^2  \qquad \text{for \ pure YM}.
  \label{eq:mD}
\end{align}
Here again, the summation is taken over species $i$ contributing to the transverse diffusion, and we use the fact that the number density is dominated by the soft population $f_\text{s} (\ell)$ for $t \gg t_\text{soft}$ and $t_\text{A} = C_\text{A}$.
For pure $\operatorname*{SU}(N)$ YM theory, $b_\alpha = - 11 N / 3$, $\mathcal{N} = 2 N \zeta (3) T_\text{s}^3 / \pi^2$, and $m_\text{D}^2 = 4 \pi \alpha (T_\text{s}) T_\text{s}^2 N / 3$.
For pure $\operatorname*{SO}(N)$ YM theory, $b_\alpha = - 22(N-2) / 3$, $\mathcal{N} = 4 (N-2) \zeta (3) T_\text{s}^3 / \pi^2$, and $m_\text{D}^2 = 8 \pi \alpha (T_\text{s}) T_\text{s}^2 (N-2) / 3$.
In the case of the SM plasma, all the SM particle contributions can be included in the soft sector to the high-energy gluon cascades such that
$b_\alpha = -7$, $\mathcal{N} = 15 \zeta(3) T^3 / \pi^2$, and $m_D^2 = 8 \pi \alpha T^2$.

Finally, we briefly discuss how the equation given in this section is related to the qualitative discussion in the previous section.
For $k \ll p$, Eq.~\eqref{eq:mu_perp} can be approximated with $\mu_\perp^2 (p; k, p-k) \sim \sqrt{ k \hat q }$.
It is consistent with the diffused transverse momentum at the formation time given in Eq.~\eqref{eq:tform}, \textit{i.e.,} $k_\perp^2 |_{t_\text{form}} \sim \hat q t_\text{form} \sim \sqrt{k \hat q}$.
The splitting function is expressed as $\gamma_{g \leftrightarrow gg} (p;k,p-k) / \nu_g \sim C_\text{A} \alpha \sqrt{ k \hat q }$ for $k \ll p$.
The corresponding splitting rate is then given as $\Gamma_\text{LPM} (k) \sim \gamma_{g \leftrightarrow gg} (p;k,p-k) / (\nu_g k) \sim C_\text{A} \alpha \sqrt{\hat q / k}$, which is consistent with Eq.~\eqref{eq:splitting_qual}.

\section{Numerical simulations}
\label{sec:numerical}

\subsection{Numerical method}

We numerically solve Eq.~\eqref{eq:Boltzmann0} without the $\mathcal{C}_{2 \leftrightarrow 2}$ term, under the assumption that the splittees of momentum $k$ with $k < k_{\rm IR}$ are thermalized immediately via $2 \to 2$ elastic scattering process.
The comoving temperature of the plasma increases
as the energy is injected into the modes $k<k_{\rm IR}$.
The physical temperature is calculated as follows:
\begin{equation}
  T_\text{s} (t)  =
  \left[ \frac{ 30}{g_\star \pi^2}
  \qty( \rho_\text{r,\,tot}(t) - \rho (t) ) \right]^{1/4},
\end{equation}
where $g_\star$ is the relativistic degrees of freedom in the thermal plasma.
We use $g_\star = 106.75$ for the SM sector and $g_\star = \nu_g$ for the pure YM theory.
The total energy injected into the radiation, $\rho_\text{r,\,tot}(t)$, is given by
\begin{align}
  \rho_\text{r,\,tot} (t) &=
  \int^t \dd t' 2 p_0 \Gamma_I n_I(t') \qty[ \frac{a (t')}{a(t)} ]^4
  \\
  &= \frac{6}{5} p_0 \Gamma_I n_I (t_0) t
  \qty[ \frac{a(t_0)}{a(t)}  ]^3 .
\end{align}
We define the energy density for non-thermal contributions as
\begin{align}
\rho (t)
&=  \nu_g \int_{k_\text{IR}} \frac{\dd^3 p}{(2\pi)^3} p f(p,t) .
\end{align}
We take $k_\text{IR} = 3 T_\text{s} (t)$ in our numerical simulations.
The results do not change qualitatively when $k_\text{IR}$ is varied by a factor of $\mathcal{O}(1)$.
We rescale the dimensionful parameters and use $T_\text{s} (t_0) \equiv T_0 = 1$ in numerical simulations without any loss of generality.

The source term $\mathcal{S}$ in Eq.~\eqref{eq:Boltzmann0} can be represented by the initial distribution \eqref{eq:initialf}.
Because the numerical time is limited, our numerical calculation starts from a finite time scale $t_0$.
This implies that the distribution at a small momentum scale is deformed as the second line of Eq.~\eqref{eq:distribution1} ($\propto T^{-7/2}$),%
\footnote{
We neglect the deformation of the distribution of the soft particles for $k > k_{\rm form}$ ($\propto k^{-3}$) for simplicity.
}
whereas the hard mode at a high momentum remains in the form of Eq.~\eqref{eq:initialf} ($\propto T^{-3/2}$).
To include this fact,
we adopt the following algorithm in our numerical simulations.
We first take an ansatz $f_\text{h}(p,t_0) = f_\text{h}(p_0) ( p/p_0 )^{-3/2}$ for the initial distribution and evolve it by a first time step, $t = t_0 +\delta t$.
Then we determine $p_\text{th,0}$ as a minimal value of momentum satisfying $\ln \left[ f_\text{h}(p,t_0+\delta t) / f_\text{h}(p,t_0) \right] < 0.1$.
Then we perform the main numerical simulations by replacing 
the initial distribution $f_\text{h} (p)$ such that
\begin{equation}
  f_\text{h}(p,t_0) =
  \begin{cases}
  f_\text{h}(p_\text{th,0})
  \qty( \frac{p}{p_\text{th,0}} )^{-7/2} &\text{for} \ \ p < p_\text{th,0} ,
  \vspace{0.2cm}
  \\
  f_\text{h}(p_0)
  \qty( \frac{p}{p_0} )^{-3/2} &\text{for} \ \ t \ge p_\text{th,0} .
  \end{cases}
  \label{eq:initialfnum}
\end{equation}
This initial condition allows the distribution function not to change significantly for the first time step, which is important to ensure the stability of numerical simulations.

The splitting term $\mathcal{C}_{1 \leftrightarrow 2}$ is given by \eqref{eq:C12}.
We take an IR cutoff for the splitting term as the initial temperature of the system $T_0$ ($=1$), assuming that temperature does not change by many orders of magnitude during the simulation time scale.
It is convenient to use a comoving momentum $\tilde{p} \equiv a(t)p/a_0$
and temperature $\tilde{T}_\text{s}(t) \equiv a(t) T_\text{s}(t) / a_0$ in numerical simulations.
We denote
\begin{equation}
  \tilde{p}_{\rm max}(t) \equiv \frac{a(t)}{a(t_0)} p_0,
\end{equation}
as the comoving momentum injected from the heavy particle decay.
To represent the IR cutoff,
we discretize the comoving momentum such that
$\Delta \tilde{p} = T_0$ ($= 1$).
We denote $i = (1,2, \dots, N_\text{grid})$ and $\tilde{p}_i = i$ as the label for the momentum grid, with $N_{\rm grid}$ being its total number.
We define $N_0$ such that $p_0 = N_0$.

The time step $\delta t$ is taken such that
$\tilde{p}_{\rm max}(t + \delta t) - \tilde{p}_{\rm max} (t) = 1$.
Namely, a single grid for comoving momentum is added at $\tilde{p}_\text{max}(t+\delta t)$ by a single time step.
Additionally, we denote $n = (0, 1,2, \dots, N_\text{grid} - N_0)$ and
\begin{equation}
  t_n = t_0 \qty( \frac{N_0 + n}{N_0} )^{3/2},
\end{equation}
for the time steps, where we use $\tilde{p}_\text{max}(t_n) = (N_0 + n)$.
For every time step, the splitting function is calculated and the distribution function is evolved as
\begin{align}
  \ln f_{n+1} (\tilde{p}_i) - \ln f_n (\tilde{p}_i) &=
  \qty( t_{n+1} - t_n )
  \qty( \frac{t_0}{t_n} )^{2/3}
  \qty( \frac{\tilde{T}_s(t_n)}{T_0} )^{3/2}
  \frac{(2\pi)^3}{\nu_g \tilde{p}_i^2} \Delta \tilde{p}
  \nonumber
  \\
  &\qquad \times
  \left[ -
  \sum_{k=1}^{i - 1} \,
    \gamma_{g \leftrightarrow g g} \bigl(\tilde{p}_i; \tilde{p}_k, \tilde{p}_{i-k} \bigr)
    +
  \sum_{k=1}^{N_0+n-i} \,
    2 \gamma_{g \leftrightarrow gg} \bigl(\tilde{p}_{i+k}; \tilde{p}_i, \tilde{p}_k \bigr) \frac{f_n (\tilde{p}_{i +k})}{f_n (\tilde{p}_{i})}
  \right]_{T \to T_0} \,,
\end{align}
where $(t_0/t_n)^{2/3}$ originates from the redshift. Here, we factorize the temperature dependence for $\gamma_{g \leftrightarrow gg}$ as $(T_s(t)/T_0)^{3/2} \times ( \gamma_{g \leftrightarrow gg} )_{T \to T_0}$ by neglecting its logarithmic dependence.

\subsection{Results of the SM}

To describe the thermalization in the SM sector, we consider the thermalization of gluon (\textit{i.e.}, $\operatorname*{SU}(3)$ gauge field) into the SM thermal plasma. Namely, we take $g_\star = 106.75$, ${\rm G} = {\rm SU}(3)$, $\nu_g = 16$, $d_A = 8$, $C_A = 3$, $b_\alpha = -7$, $\mathcal{N} = 15 \zeta(3) T^3/\pi^2$, $m_D^2 = 8 \pi \alpha T^2$, and $\alpha(m_Z) = 0.118$ with $m_Z$ being the $Z$-boson mass. We take $T_0 = 10^3 \GeV = 1$ in the numerical simulations, where the dependence on this choice is only because of the renormalization group (RG) running of the gauge coupling constant.

In this case, there remains three parameters that should be specified to perform the numerical simulations:
\begin{align}
  f_\text{h}(p_0) \ ({\rm or} \ \Gamma_I), \quad t_0, \quad p_0\,.
\end{align}
Here,
$f_\text{h}(p_0)$ should be consistently determined by the first line of \eqref{eq:Ts} with $t = t_0$. This implies that $f_\text{h}(p_0) \propto \nu_g^{-1} g_\star \alpha^{-4} t_0^{-2} p_0^{-3}$.
We are interested in the regime around $t \sim t_\text{max}$;
thus, we take $t_0$ to be of order but smaller than $t_\text{max}$.
Moreover, we need to ensure $t_{\rm soft} \ll t_0$.
Specifically,
we take
\begin{equation}
  t_0 \sim
  \epsilon_t \frac{p_0^{1/2}}{\alpha^2 T_0^{3/2}},
\end{equation}
with
\begin{equation}
  \sqrt{\frac{T_0}{p_0}} \ll \epsilon_t \lesssim 1,
  \label{eq:epsilon_t}
\end{equation}
from \eqref{eq:ksplit}, where we use \eqref{eq:Ts} to eliminate $\tilde{\Gamma}$ or $f_\text{h}(p_0)$.
The lower bound on $\epsilon_t$ originates from the condition of $t_{\rm soft} \ll t_0$.
In the numerical simulations,
we primarily take
\begin{align}
  &p_0 =
  5 \times 10^3\,,
  \\
  &t_0 =
  2 \times p_0^{1/2}\,,
  \\
  &f_\text{h}(p_0)
  = 10^5 \times
  t_0^{-2} p_0^{-3}\,,
  \label{eq:initial}
\end{align}
and $N_{\rm grid} = 2 \times 10^4$.
We also perform numerical calculations by changing $t_0$ and $p_0$ by a factor of a few to determine whether the result depends only on physical parameters.

\begin{figure}
    \centering
            \includegraphics[width=0.55\hsize]{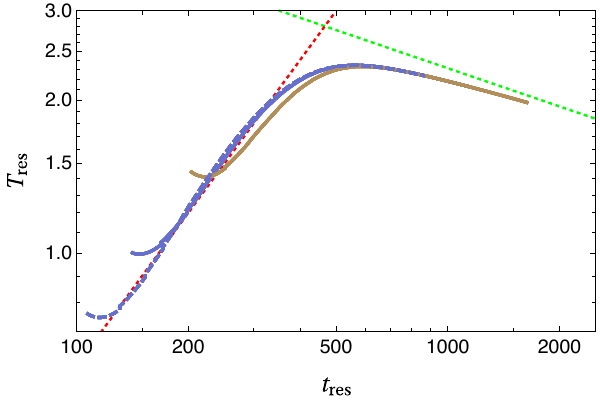}
    \caption{
    Physical temperature as a function of physical time. All curves are rescaled according to \eqref{eq:rescale1} and \eqref{eq:rescale2}.
    The red and green dotted lines represent the analytic dependence for $T_\text{s}(t)$ in the regimes of $t \ll t_{\rm max}$ ($T_\text{s}(t) \propto t$) and $t \gg t_{\rm max}$ ($T_\text{s}(t) \propto t^{-1/4}$), respectively.
    The blue solid/dashed curves represent the numerical results for $( p_0, t_0 p_0^{-1/2}, f_\text{h} (p_0) t_0^2 p_0^3 ) = (5 \times 10^3,2,10^5), (10^4, 1, 10^5)$ respectively.
    The brown solid curve corresponds to $(5 \times 10^3, 5, 10^5)$.
}
    \label{fig:result1}
\end{figure}

Figure~\ref{fig:result1} shows the temperature as a function of time, where the variables are rescaled by
\begin{align}
& T_s(t)  = T_\text{res} (t)
\qty( \frac{f_\text{h}(p_0)}{10^5\, t_0^{-2} p_0^{-3}} )^{2/5}
\qty( \frac{t_0}{2 p_0^{1/2}} )^{-2/5} \,,
\label{eq:rescale1}
\\
& t_n  = t_\text{res} \qty( \frac{f_\text{h}(p_0)}{10^5 \, t_0^{-2} p_0^{-3}} )^{-3/5} \qty( \frac{t_0}{2 p_0^{1/2}} )^{3/5} \qty( \frac{p_0}{5 \times 10^3} )^{1/2}.
\label{eq:rescale2}
\end{align}
These dependencies on the initial parameters are expected from the analytic estimations of Eqs.~\eqref{eq:Tmax}, \eqref{eq:tthermal}, and Eq.~\eqref{eq:gamma-to-fh}.
The blue solid curve represents the case with the initial condition mentioned above.
The brown solid curve represents the case with $p_0 = 5 \times 10^3$, $t_0 = 5 p_0^{1/2}$, $f_\text{h}(p_0) = 10^5 \, t_0^{-2} p_0^{-3}$, and $N_{\rm grid} = 2 \times 10^4$.
The dashed curve represents the case with $p_0 = 10^4$, $t_0 = p_0^{1/2}$, $f_\text{h}(p_0) = 10^5 \, t_0^{-2} p_0^{-3}$, and $N_{\rm grid} = 4 \times 10^4$.
All results are in good agreement with each other, except for the regime around $t=t_0$. The discrepancy results from the fact that the initial distribution (and $f_\text{h}(p_0)$) is chosen by hand for soft modes. However, because the energy injection into thermal plasma is dominated by the contribution from hard modes, the late-time distribution is not affected, \textit{i.e.,} implying the basin of attraction.
The results for other initial conditions are provided in the Appendix.

The red and green dotted lines represent the fitting functions for $T(t)$:
\begin{equation}
  T_\text{res} (t) \simeq
  \begin{cases}
  0.006 \times t_{\rm res} & \text{for} \quad t \ll t_{\rm max} \,,
  \vspace{0.2cm}
  \\
  13 \times t_{\rm res}^{-1/4} & \text{for} \quad t \gg t_{\rm max} \,.
  \end{cases}
  \label{eq:fit}
\end{equation}
All results are in good agreement with the analytic estimations.
The asymptotic behavior for $t \gg t_\text{max}$ is further confirmed by simulations with different initial conditions as we show in Fig.~\ref{fig:initial} in Appendix. 
From Fig.~\ref{fig:result1}, the numerical coefficients of the maximal temperature and corresponding time can be determined as follows:
\begin{align}
T_\text{max} &\simeq 2.3 \times
\qty( \frac{f_\text{h}(p_0)}{10^5 t_0^{-2} p_0^{-3}} )^{2/5}
\qty( \frac{t_0}{2 p_0^{1/2}} )^{-2/5} \,,
\nonumber
\\
&\simeq 0.050 \times \qty(\frac{\Gamma_I}{m_I^3 / \Mpl^2})^{2/5} m_I,
\label{eq:resultTmax}
\\
t_\text{max} &\simeq 5.0 \times 10^2
\qty( \frac{f_\text{h}(p_0)}{10^5 t_0^{-2} p_0^{-3}} )^{-3/5} \qty( \frac{t_0}{2 p_0^{1/2}} )^{3/5} \qty( \frac{p_0}{5 \times 10^3} )^{1/2} \,,
\nonumber
\\
&\simeq 1.6 \times 10^3 \qty(\frac{\Gamma_I}{m_I^3 / \Mpl^2})^{-3/5} m_I^{-1}.
\label{eq:resulttmax}
\end{align}
These results confirm the analytic estimations \eqref{eq:Tmax} and \eqref{eq:tthermal}, and the numerical prefactors are determined from our numerical simulations.
Note that $\alpha$ and $g_\star$ are absorbed into the numerical factors because we substitute the SM values for them.

\begin{figure}
    \centering
            \includegraphics[width=0.45\hsize]{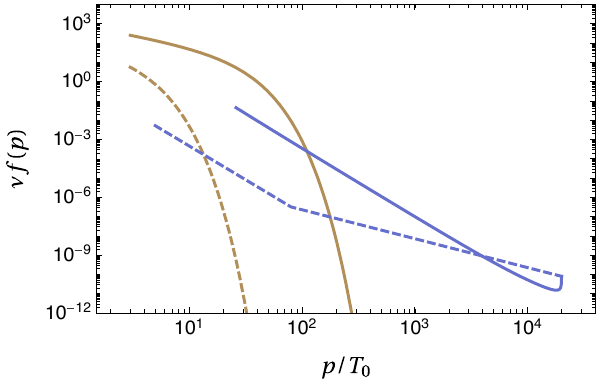}
            \hspace{0.4cm}
            \includegraphics[width=0.45\hsize]{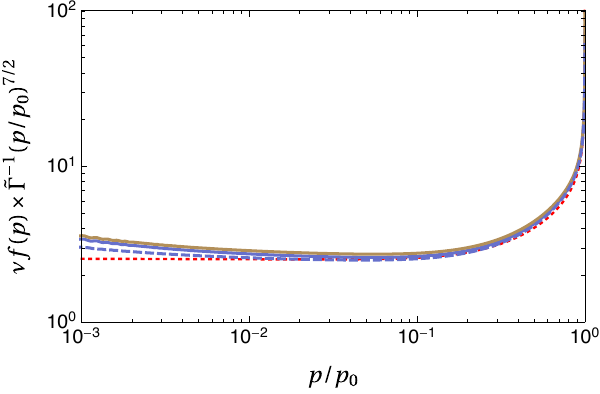}
    \caption{
    Spectra at the end of numerical simulations.
    (left) The blue/brown solid curves represent the hard/thermal spectra respectively. We also show the initial spectra with the dashed lines in the same color.
    (right) The red dotted curve represents the stationary solution corresponding to $\Gamma_{\rm split}/ H \to 0$ given in Eq.~\eqref{eq:stationary}.
    The color codings for the blue solid/dashed and brown curves are the same as Fig.~\ref{fig:result1}.
    }
    \label{fig:result2}
\end{figure}

The blue curves in Fig.~\ref{fig:result2} show the spectra at the end of numerical simulations.
In the left panel, the initial spectrum \eqref{eq:initialfnum} is represented by the blue dashed curve.
The solid (dashed) brown curve represents the thermal spectrum at the end (beginning) of numerical simulations.
See Eqs.~\eqref{eq:distribution1} and \eqref{eq:distribution2} for analytic estimations.
In the right panel, the distributions are rescaled by $\tilde{\Gamma}(p/p_0)^{-7/2}$, where
\begin{equation}
  \tilde{\Gamma} \equiv \frac{4 \pi^2 \Gamma_I n_I(t)}{p_0^{5/2} T_\text{s}^{3/2} (t) }
  = f_\text{h}(p_0) \nu_g \frac{H_0 p_0^{1/2}}{\tilde T_\text{s}^{3/2} (t)} \, \frac{t_0}{t},
\end{equation}
is defined in our previous paper~\cite{Mukaida:2022bbo}.
The three curves correspond to the ones used in Fig.~\ref{fig:result1} and almost overlap with each other.
At a late time, $\Gamma_\text{split} \gg H$, in which case
the spectrum can be represented by a stationary solution of
\begin{equation}
  \label{eq:stationary}
  \nu_g f(p,t) \simeq
  2.6 \times
  \tilde{\Gamma} \left(\frac{p}{p_0} \right)^{-7/2},
\end{equation}
for $p \ll p_0$.
The red dotted curve, which is also overlapped with our numerical results, is the result for the stationary solution of the Boltzmann equation in the limit of $\Gamma_\text{split} \gg H$.
All results are consistent with the stationary solution, which justifies $t \gg t_\text{max}$ at the end of numerical simulations.
The deviation from the dependence of $\propto p^{-7/2}$ at a small $p/p_0$ may be due to the IR cutoff used in our numerical simulations.
For the stationary solution, the IR cutoff need not to be introduced, as explained in Ref~\cite{Harigaya:2014waa,Mukaida:2022bbo}; therefore, the stationary solution is almost exactly $\propto p^{-7/2}$ for a small $p/p_0$.

\subsection{Results of pure YM theory}

Now we consider a pure YM theory for ${\rm G = SU}(N)$ and $\operatorname*{SO}(N)$
with $(N,\alpha) = (3, 10^{-1})$, $(5, 10^{-2})$,
$(5, 10^{-3})$,
$(5, 10^{-4})$, and $(10, 10^{-3})$ in a dark sector,
where the gauge coupling constant is defined at the energy scale of $m_Z$. We take $T_0 = 10^3 \GeV$ ($\equiv 1$) as an example, though our results depends on its value only logarithmically through the RG running.

In the numerical simulations, we take
\begin{align}
  &p_0 =
  p_0^{\rm (bm)} \equiv
  5 \times 10^3,
  \\
  &t_0 =
  t_0^{\rm (bm)} \equiv 2 \times p_0^{1/2} \left( \frac{\Gamma_{\rm split} (p_0)}{\Gamma_{\rm split, SM}(p_0)} \right)^{-1},
  \label{eq:t_02}
  \\
  &f_\text{h}(p_0) = f_\text{h}^{\rm (bm)} \equiv 10^5 \times
  t_0^{-2} p_0^{-3}
  \left( \frac{\nu_g}{\nu_{\rm SM}} \right)^{-1}
  \left( \frac{g_*}{g_{*, {\rm SM}}} \right)
  \left( \frac{\Gamma_{\rm split}(p_0)}{\Gamma_{\rm split, SM}(p_0)} \right)^{-2},
\end{align}
and $N_{\rm grid} = 2 \times 10^4$,
for $(N,\alpha) = (3, 0.1)$,
where $\nu_{\rm SM} = 16$ and $g_{*, {\rm SM}} = 106.75$.
Here, we define
\begin{equation}
  \Gamma_{\rm split}(p_0) \equiv \frac{1}{\nu_g p_0} \gamma_{g \leftrightarrow gg} (p_0;p_0/2,p_0/2).
\end{equation}

For the case with $\alpha \sim 0.1$, including the SM QCD,
the difference of the gauge coupling constants at $p_0$ and at $T_0$ due to the RG running is not negligible. For example,
\begin{equation}
  \frac{\alpha(p_0)}{\alpha (T_0)} \simeq 0.5,
\end{equation}
for $(N,\alpha) = (3,10^{-1})$ with ${\rm G = SU}(N)$.
In this case, $\alpha$ in $t_{\rm soft}$ is relatively larger than that used in $t_{\rm max}$.
This results in a relatively weaker condition on the initial time, \eqref{eq:epsilon_t}, by a factor of a few.
In contrast, for a significantly smaller gauge coupling constants, the RG running is negligible.
In this case, the lower bound on the initial time, given by the first inequality of \eqref{eq:epsilon_t},
is more severe than the case of the SM.
Therefore, we instead take $t_0 = 5 \times t_0^{\rm (bm)}$ for
$(N,\alpha) = (5, 10^{-2}), (5, 10^{-3}), (5, 10^{-4})$, and $(10, 10^{-3})$.

\begin{figure}
    \centering
            \includegraphics[width=0.45\hsize]{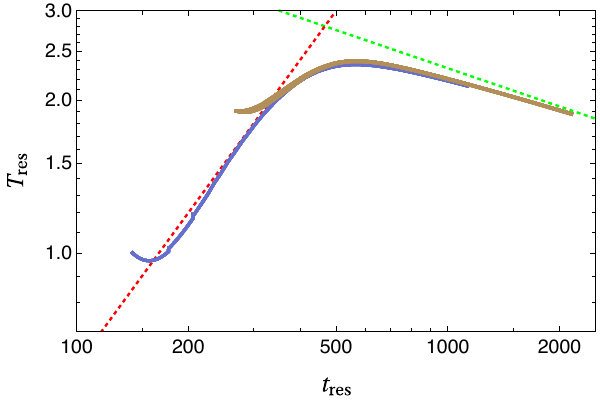}
            \hspace{0.4cm}
            \includegraphics[width=0.45\hsize]{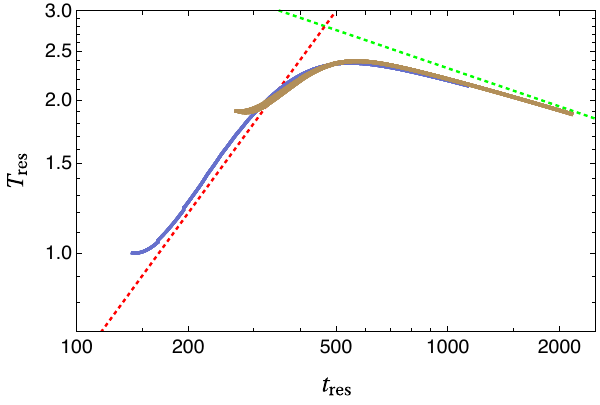}
    \caption{
    Same as Fig.~\ref{fig:result1} but for the pure $\operatorname*{SU}N$ (left panel) and $\operatorname*{SO}(N)$ (right panel) YM theories.
    The blue curve represents the result of $(N,\alpha) = (3, 10^{-1})$, whereas the brown curves represent the results of $(5, 10^{-2})$, $(5, 10^{-3})$, $(5, 10^{-4})$, and $(10, 10^{-3})$.
}
    \label{fig:resultYM}
\end{figure}

Figure~\ref{fig:resultYM} shows the temperature as a function of time in pure YM theories.
To show our results, we rescale parameters such as
\begin{align}
& T(t)  = T_\text{res} (t)
\qty( \frac{f_\text{h}(p_0)}{f_\text{h}^{\rm (bm)}} )^{2/5}
\qty( \frac{t_0}{t_0^{\rm (bm)}} )^{-2/5} \,,
\label{eq:rescale3}
\\
& t_n  = t_\text{res} \qty( \frac{f_\text{h}(p_0)}{f_\text{h}^{\rm (bm)}} )^{-3/5} \qty( \frac{t_0}{t_0^{\rm (bm)}} )^{3/5} \qty( \frac{p_0}{p_0^{\rm (bm)}} )^{1/2}
\left( \frac{\Gamma_{\rm split} (p_0)}{\Gamma_{\rm split, SM}(p_0)} \right)^{-1}.
\label{eq:rescale4}
\end{align}
We plot the results of $\operatorname*{SU}(N)$ (left panel) and $\operatorname*{SO}(N)$ (right panel) gauge theory with $(N,\alpha) = (3, 10^{-1})$ as blue solid curve and those with $(5, 10^{-2})$, $(5, 10^{-3})$, $(5, 10^{-4})$, and $(10, 10^{-3})$ as brown solid curves.
The brown solid curves overlap and cannot be distinguished from each other.
The dotted lines are the fitting functions of \eqref{eq:fit}.
The figure shows that all results are consistent with \eqref{eq:fit} after the rescalings.
Moreover,
the results of \eqref{eq:resultTmax} and \eqref{eq:resulttmax} can be applied to pure YM theories
after the correction from the difference of the splitting rate, $\Gamma_{\rm split}(p_0)$, is included:
\begin{align}
  T_\text{max} &\simeq 2.3 \times
  \qty( \frac{f_\text{h}(p_0)}{f_\text{h}^{\rm (bm)}} )^{2/5}
  \qty( \frac{t_0}{t_0^{\rm (bm)}} )^{-2/5}
  \nonumber
  \\
  &\simeq 2.2 \times
  g_*^{-2/5}
  \left( \frac{\Gamma_{\rm split} (p_0)}{T_\text{s}^{3/2} p_0^{-1/2}} \right)^{2/5}
  \qty(\frac{\Gamma_I}{m_I^3 / \Mpl^2})^{2/5} m_I,
  \label{eq:resultTmaxYM}
  \\
  t_\text{max} &\simeq 5.0 \times 10^2  \times
  \qty( \frac{f_\text{h}(p_0)}{f_\text{h}^{\rm (bm)}} )^{-3/5} \qty( \frac{t_0}{t_0^{\rm (bm)}} )^{3/5} \qty( \frac{p_0}{p_0^{\rm (bm)}} )^{1/2}
  \left( \frac{\Gamma_{\rm split} (p_0)}{\Gamma_{\rm split, SM}(p_0)} \right)^{-1}
  \nonumber
  \\
  &\simeq 3.0 \times 10^{-3}
  g_*^{3/5}
  \left( \frac{\Gamma_{\rm split} (p_0)}{T_\text{s}^{3/2} p_0^{-1/2}} \right)^{-8/5}
  \qty(\frac{\Gamma_I}{m_I^3 / \Mpl^2})^{-3/5} m_I^{-1}\,,
\label{eq:resulttmaxYM}
\end{align}
where we substitute the SM parameters, such as $\Gamma_{\rm split}(p_0) \simeq 5.5 \times 10^{-4} \times T_\text{s}^{3/2} p_0^{-1/2}$, and
$p_0 = m_I/2$.
This is a general result that should be applicable to any pure YM(-like) theories.

We also provide useful formulas for $T_{\rm max}$ and $t_{\rm max}$ in large $N$ and small $\alpha$ limits.
If $N \gg 1$ and $\alpha \ll 0.1$, we can approximate
\begin{equation}
  \Gamma_{\rm split} (p_0) \approx
  \left\{
  \begin{array}{ll}
  (0.01\,\text{-}\,0.02) \times N^2 \alpha^2 T^{3/2}p_0^{-1/2} \qquad \text{for} \quad \operatorname*{G} = \operatorname*{SU}(N) ,
  \\
  (0.03\,\text{-}\,0.07) \times N^2 \alpha^2 T^{3/2}p_0^{-1/2} \qquad \text{for} \quad \operatorname*{G} = \operatorname*{SO}(N) ,
  \end{array}
  \right.
\end{equation}
where the prefactors depend on $p_0$ logarithmically.
Substituting these into \eqref{eq:resultTmaxYM} and \eqref{eq:resulttmaxYM}, we obtain
\begin{align}
  T_\text{max} &\approx (0.3\,\text{-}\,0.4) \times
  \alpha^{4/5}
  \qty(\frac{\Gamma_I}{m_I^3 / \Mpl^2})^{2/5} m_I
  \,,
  \label{eq:resultTmaLargeSUN}
  \\
  t_\text{max} &\approx (2\,\text{-}\,5)
  \times
  N^{-2} \alpha^{-16/5}
  \qty(\frac{\Gamma_I}{m_I^3 / \Mpl^2})^{-3/5} m_I^{-1} \,,
  \label{eq:resulttmaxLargeSUN}
\end{align}
for $\operatorname*{G} = \operatorname*{SU}(N)$
and
\begin{align}
  T_\text{max} &\approx (0.5\,\text{-}\,0.7) \times
  \alpha^{4/5}
  \qty(\frac{\Gamma_I}{m_I^3 / \Mpl^2})^{2/5} m_I
  \,,
  \label{eq:resultTmaxLargeSON}
  \\
  t_\text{max} &\approx (0.2\,\text{-}\,0.8) \times
  N^{-2} \alpha^{-16/5}
  \qty(\frac{\Gamma_I}{m_I^3 / \Mpl^2})^{-3/5} m_I^{-1}\,,
  \label{eq:resulttmaxLargeSON}
\end{align}
for ${\rm G = SO}(N)$,
where we use $g_* = \nu_g$.
These results confirm the analytic estimations of Eqs.~\eqref{eq:tthermal} and \eqref{eq:Tmax}, where $D_s \sim C_A^2 \sim N^2$ and $g_\star = \nu_g \sim N^2$.

\section{Discussion and conclusions}
\label{sec:conclusions}

We have obtained a clear dynamical picture of the thermalization of pure YM plasma during perturbative reheating after inflation by numerically solving the Boltzmann kinetic equation by appropriately considering the LPM effect.
Our results confirm the results of previous analytic studies based on quasi-equilibrium ansatz applicable to two regimes of $t_\text{soft} \ll t \ll t_\text{max}$ or $t_\text{max} \ll t$, but also describe the transient epoch between these regimes, providing a better estimation for the maximal temperature of our Universe.
The maximal temperature obtained can be significantly smaller than the estimation based on the instantaneous thermalization approximation, highlighting the importance of understanding thermalization.
Moreover, our results demonstrate the robustness of the quasi-equilibrium solution under the consistent change of the initial conditions.
Furthermore, the late-time behavior is stable even when the initial conditions are applied far away from the consistent initial conditions.
The implications of our results for particle cosmology are broad.
In particular, the maximal temperature of the Universe is the key ingredient for understanding possible cosmological phase transitions.
Recently, hidden pure YM theories have been gaining attention because they involve the glueballs as a candidate of DM~\cite{Faraggi:2000pv,Feng:2011ik,Boddy:2014yra,Boddy:2014qxa,Soni:2016gzf,Kribs:2016cew,Forestell:2016qhc,Soni:2017nlm,Forestell:2017wov,Jo:2020ggs,Gross:2020zam,Carenza:2022pjd,Carenza:2023shd,Yamada:2023thl} and would lead to the first-order phase transitions~\cite{Reichert:2021cvs,Morgante:2022zvc,He:2022amv,Reichert:2022naa}, possibly accompanied by cosmic strings~\cite{Witten:1985fp,Yamada:2022imq,Yamada:2022aax}.
Our results are essential to understanding their implications, such as the prediction of the GW spectrum as well as the condition to obtain the confinement phase transition after inflation.
As discussed in the main text, we believe our results can be applicable to the SM plasma when the inflaton dominantly decays into the SM gluons of $\operatorname*{SU}(3)$.
This expectation is based on our previous study~\cite{Mukaida:2022bbo}, which showed that the thermalization is dominated by the gluons, although the analysis is restricted to the quasi-equilibrium regime.
The complete dynamical analysis of the SM plasma, including the cases in which the inflaton decays into other SM species, is worthwhile to investigate in future studies.

{\small
\section*{Acknowledgments}
K.\,M.\ was supported by MEXT Leading Initiative for Excellent Young Researchers Grant No.\ JPMXS0320200430, and JSPS KAKENHI Grant No.\ JP22K14044.
M.\,Y.\ was supported by MEXT Leading Initiative for Excellent Young Researchers, and by JSPS KAKENHI Grant No.\ JP20H05851 and JP23K13092.
}

\appendix

\section{Case with other initial conditions}
\label{sec:initial}

In the main part of this paper, we primarily consider the case of thermal plasma generated through the thermalization of injected high-energy particles. This requires a consistent initial condition such as \eqref{eq:initial}.
In this Appendix, we provide results with other initial conditions, particularly for the case with an SM-like system.

\begin{figure}
  \centering
    \includegraphics[width=0.55\hsize]{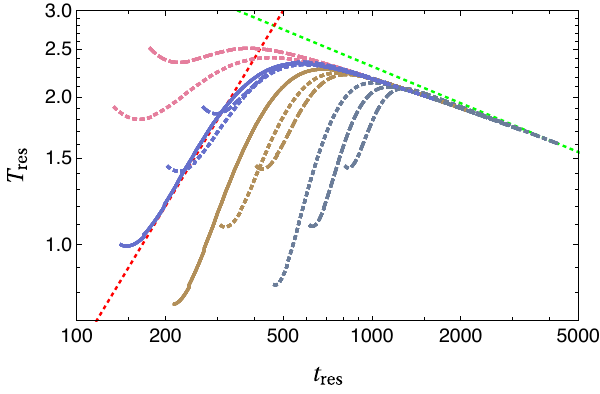}
    \caption{
    Same as Fig.~\ref{fig:result1} but with different initial conditions.
    See Eq.~\eqref{eq:various_initial_conds} for their color codings.
    }
    \label{fig:initial}
\end{figure}

Figure~\ref{fig:initial} shows the results for the cases with
\begin{equation}
(t_0/p_0^{1/2}, \, f_\text{h}(p_0) / (10^5 \times
  t_0^{-2} p_0^{-3}))
  =
  \begin{cases}
  (5,1/2), \,
  (10,1/2)   & \cdots~ \text{(magenta dotted, dashed)}
  \\
  (2,1), \,  (5,1), \,
  (10,1)   & \cdots~ \text{(blue solid, dotted, dashed)}
  \\
  (2,2), \,
  (5,2), \, (10,2)  & \cdots~ \text{(brown solid, dotted, dashed)}
  \\
  (5,4), \, (10,4), \,
  (20,4)  & \cdots~ \text{(cyan dotted, dashed, dot-dashed)}
  \end{cases}
  \label{eq:various_initial_conds}
\end{equation}
with $p_0 = 5 \times 10^3$
and $N_{\rm grid} = 1.5 \times 10^4$.
The variables are rescaled according to \eqref{eq:rescale2}.

Note that the blue dotted curve corresponds to the result of \eqref{eq:initial}.
The numerical simulations for a smaller $f_\text{h}(p_0) / (10^5 \times t_0^{-2} p_0^{-3})$
correspond to the case with a relatively higher initial temperature for the ambient plasma.
This is the case in which ambient plasma is already generated by other mechanisms or sources.
Once the energy injection from the high-energy particle is sufficiently high,
the temperature starts to increase and reaches its maximal value within the time scale of $\mathcal{O}(t_{\rm max})$.

On the contrary,
the numerical simulations for a larger $f_\text{h}(p_0) / (10^5 \times t_0^{-2} p_0^{-3})$
imply a relatively lower initial temperature for the ambient plasma.
This corresponds to cases in which the thermalization via the elastic scatterings in the soft sector
or the energy injection into the soft sector are delayed by some mechanisms until $t = t_0$.
In these cases,
the temperature increases faster to reach the expected behavior of \eqref{eq:Ts}.
However,
because the simulation time is limited such that $t_0$ and $t_{\rm max}$ are of the same order of magnitude, we cannot confirm that the temperature reaches the attractor of \eqref{eq:Ts} before it reaches the maximal temperature.
Still,
all results reach the maximal temperature
within the time scale of order $t_{\rm max}$, even if the simulation starts at a later time.
All results agree at a later time ($t \gg t_{\rm max}$), at which
high-energy particles thermalize within the Hubble-time scale.

We expect that the cases with $f_\text{h}(p_0) / (10^5 \times
t_0^{-2} p_0^{-3}) = 1$ (blue curves) are consistent initial conditions in which the ambient plasma is generated by the energy injection via the thermalization of high-energy particles. These cases agree with each other except for $t \sim t_0$, even if we change the value of $t_0/p_0^{1/2}$. This supports the fact that it consistently starts within the attractor regime of \eqref{eq:Ts}.
This is not the case for different values of $f_\text{h}(p_0) / (10^5 \times t_0^{-2} p_0^{-3})$, as shown in the figure.
However, we note that the other initial conditions may also be interesting in some of the cases mentioned above.
Our numerical simulations can also be used to analyze such cases.

\small
\bibliographystyle{utphys}
\bibliography{draft}

\end{document}